\documentclass{cjaa}           % referee version: for submission
\usepackage{graphicx}                   %for PS/EPS graphics inclusion, new
\input{epsf.sty}                        %for PS/EPS graphics inclusion, old
\input{psfig.sty}                       %for PS/EPS graphics inclusion, old

\begin{document}

   \title{Neutrinos from microquasars}

   \volnopage{Vol.0 (200x) No.0, 000--000}      %%preserved for Editor. DOn't remove!
   \setcounter{page}{1}          %%starting page, preserved for Editor. DOn't remove!

   \author{Diego F. Torres
      \inst{1}
%% Please move "\mailto{}" to the corresponding author of the paper.
   \and Gustavo E. Romero
      \inst{2,3}
   \and Felix Mirabel
      \inst{4,5}
      }
   \offprints{D. F. Torres}                   %% is disabled in fact

   \institute{Lawrence Livermore National Laboratory, 7000 East Avenue, L-413,
         Livermore, CA 94550. Email: {dtorres@igpp.ucllnl.org}
%% offprint requests will be sent. Note to pair \mailto{} with \email{}
        \and
        Facultad de Ciencias Astron\'omicas y Geof\'{\i}sicas,
        Universidad Nacional de La Plata, CC 1900, La Plata,
        Argentina
        \and
             Instituto Argentino de Radioastronomía, C.C.5, (1894) Villa Elisa,
         Buenos Aires, Argentina
        \and
             CEA/DSM/DAPNIA/Service d'Astrophysique, Centre
d'Etudes de Saclay, F-91191 Gif-sur-Yvette, France \and Instituto de
Astronom\'{\i}a y F\'{\i}sica del Espacio/CONICET, C.C. 67, Suc. 28,
Buenos Aires, Argentina
          }

   \date{}

   \abstract{The jets of microquasars with high-mass stellar
companions are exposed to the dense matter field of the stellar wind
as well as to the photon densities found in the surrounding medium.
Photopion and proton-proton interactions could then lead to copious
production of neutrinos. In this work, we analyze the hadronic
microquasar model, particularly in what concerns to the neutrino
production. Limits to this kind of models using data from AMANDA-II
are established. New constraints are also imposed upon specific
microquasar models based on photopion processes. These are very
restrictive particularly for the case of SS433, a microquasar for
which the presence of accelerated hadrons has been already inferred
from iron X-ray line observations.
   }

   \authorrunning{Torres, Romero, \& Mirabel}
%author_head in even pages
   \titlerunning{Neutrinos from microquasars}
% title_head in odd pages

   \maketitle
%% Note: In the following text body of your manuscript, please note several differences from
%%       other major journals:
%% (1) \subsection{Please Capitalize the First Letter of Each Notional Word in Subsection Title}
%% (2) Please Capitalize the First Letter of Each Notional Word in table's caption

%
%________________________________________________ sections below
%
\section{Introduction}           %% first-level sections will be auto-capitalized
\label{sect:intro}
%\hspace{15pt}%                   %% preserved for Editor

The presence of relativistic hadrons in microquasar jets like those
of SS433 has been already inferred from iron X-ray line observations
(e.g. Migliari et al. 2002). A significant content of relativistic
hadrons in microquasar jets could open the possibility for a
hadronically-generated emission of high energy radiation. Here, we
discuss a new mechanism for the generation of high-energy gamma-rays
in microquasars that is based on hadronic interactions occurring
outside the coronal region (Romero et al. 2004).  In this model the
gamma-ray and neutrino emission arises from the decay of neutral
pions created in the inelastic collisions between relativistic
protons ejected by the compact object and  ions in the stellar wind.
The only requisites for the model to operate are a windy high-mass
stellar companion and the presence of multi-TeV protons in the jet,
both of which seem natural in microquasar environments. We pay
particular attention to the possible neutrino signal of this kind of
models, and impose constraints using the latest AMANDA-II data
(Ahrens et al. 2004).

\section{The hadronic microquasar model}
\label{sect:Thevar}
%\hspace{15pt}%                   %% preserved for Editor

\subsection{The jet and the particle spectrum in the lab frame}

For simplicity, we shall not make any specific assumption about the
magnetic field or other parameters in the jet, but rather model it
as a beam of energetic particles (i.e., in the spirit of, for
instance, Purmohammad and Samimi 2001, and Bednarek et al. 1990).
The jet axis, $z$, is assumed normal to the orbital radius $a$. We
shall allow the jet to expand laterally, in such a way that its
radius is given by $R(z)=\xi z^{\epsilon}$, with $\epsilon\leq 1$
and $z_0\leq z\leq z_{\rm max}$. For $\epsilon=1$ we have a conical
beam. The jet starts to expand at a height $z_0\sim$ a few hundred
km above the black hole, outside the coronal region. The particle
spectrum of the relativistic $e-p$ flow is assumed to be a power law
$N'_{e,\;p}(E'_{e, \;p})= K_{e, \;p}\; {E'}_{e, \;p}^{-\alpha}$,
valid for $ {E'_{e, \;p}}^{\rm min}\leq  E'_{e, \;p} \leq  {E'_{e,
\;p}}^{\rm max}$, in the jet frame. The corresponding particle flux
will be $J'_{e, \;p}( E'_{e, \;p})= (c/4\pi)
N'_{e,\;p}(E'_{e,\;p})$. Since the jet expands, the proton flux can
be written as:
\begin{equation}
J'_p(E'_p)=\frac{c}{4 \pi} K_0 \left(\frac{z_0}{z}\right)^{\epsilon
n} {E'_p}^{-\alpha}, \label{Jp}
\end{equation}
where $n>0$ (a value $n=2$ corresponds to the conservation of the
number of particles, see Ghisellini et al. 1985), and a prime refers
to the jet frame.
Note that these expressions are valid in the jet frame.
Using relativistic invariants, it can be proven
that the proton flux, in the observer (or lab) frame, becomes (e.g.
Purmohammmad \& Samimi 2001)
\begin{equation}
J_p(E_p,\theta)=\frac{c K_0}{4 \pi} \left(\frac{z_0}{z}\right)^
{\epsilon n} \frac{\Gamma^{-\alpha+1} \left(E_p-\beta_{\rm b}
\sqrt{E_p^2-m_p^2c^4} \cos \theta\right)^{-\alpha}}{\left[\sin ^2
\theta + \Gamma^2 \left( \cos \theta - \frac{\beta_{\rm b}
E_p}{\sqrt{E_p^2-m_p^2 c^4}}\right)\right]^{1/2}}, \label{Jp_lab}
\end{equation}
where $\Gamma$ is the jet Lorentz factor, $\theta$ is the angle
subtended by the emerging photon direction (assumed to be similar to
the initial proton direction) and the jet axis, $\beta_{\rm b}$ is
the corresponding velocity in units of $c$. Note that only photons
emitted with angles similar to that of the inclination angle of the
jet will reach a distant observer, and thus $\theta$ can be
approximated by the jet inclination angle.

In order to obtain Eq. (\ref{Jp_lab}) one has to consider
conservation of particles. If $n$ is the number of protons per unit
energy per unit solid angle per unit volume, so that in the frame
comoving with the jet one has $n^\prime(E^\prime,\Omega^\prime)=
(A/4\pi) E^{\prime -\alpha} dV^\prime$, the equality $n(E,\Omega) dV
d\Omega dE=n^\prime(E^\prime,\Omega^\prime) dV^\prime d\Omega^\prime
dE^\prime$ holds, what implies
$n(E,\Omega)=n^\prime(E^\prime,\Omega^\prime)
(d\Omega^\prime/d\Omega)(dE^\prime/dE) (dV^\prime/dV)$. From the
invariance of $d^3p/E$, the invariance of $p d\Omega dE$ is also
proved (Hayakawa 1969, p.715), so that
$(d\Omega/d\Omega^\prime)(dE^\prime/dE)=p/p^\prime$. Using
relativistic Lorentz transformations, $p_\| ^\prime = \Gamma (p_\|
+\beta E)$, with $p_\|=p \cos \theta$ and $p_\perp ^\prime =
p_\perp=p \sin \theta$, what implies $p^{\prime 2} = p^2 ( \sin^2
\theta + \Gamma^2 (\cos \theta + \beta E / \sqrt{E^2-m^2})^2)$. This
defines $p/p^\prime$, which, together with the equalities
$dV^\prime/dV=\Gamma$ and $E^\prime = \Gamma (E-\beta \sqrt{E^2-m^2}
\cos \theta)$, yields to Eq. (\ref{Jp_lab}).

We will adopt the jet-disk coupling hypothesis proposed by Falcke \&
Biermann (1995) and applied with success to AGNs, i.e. the total jet
power scales with the accreting rate as $Q_{\rm j}=q_{\rm j}
\dot{M}_{\rm disk} c^2$, with $q_{\rm j}=10^{-1}-10^{-3}$. The
number density ${n_0}'$ of particles flowing in the jet at
$R_0=R(z_0)$ is then given by $c\pi R_0^2 {n_0}'=Q_{\rm j}/m_p c^2$,
where $m_p$ is the proton rest mass. This implies:
\begin{equation}
{n_0}'=\frac{q_{\rm j} \dot{M}_{\rm disk}}{\pi c  R_0^2 m_p}
\label{n01}
\end{equation}
Additionally, $n_0'=\int^{{E'}_p^{\rm max}}_{{E'}_p^{\rm min}}
{N'}_p({E'}_p, \;z_0)\; d{E_p}'$. Then, if ${E'}_p^{\rm
max}>>{E'}_p^{\rm min}$, which is always the case, we have
\begin{equation}
K_0={n_0}' (\alpha-1) ({E'}_p^{\rm min})^{\alpha-1},\label{n02}
\end{equation}
which gives the constant in the power-law spectrum at $z_0$. This
completely defines the proton spectrum.

\subsection{Simple wind modelling}

The structure of the matter field in the wind will be determined
essentially by the stellar mass loss rate and the continuity
equation: $\dot{M_*}=4\pi r^2 \rho (r) v(r)$, where $\rho$ is the
density of the wind and $v$ is its velocity. Hence,
\begin{equation}
\rho(r)=\frac{\dot{M_*}}{4\pi r^2 v(r)}.
\end{equation}
The radial dependence of the wind velocity is given by (Lamers \&
Cassinelli 1999):
\begin{equation}
v(r)=v_{\infty}\left(1-\frac{r_*}{r}\right)^{\beta},
\end{equation}
where $v_{\infty}$ is the terminal wind velocity, $r_*$ is the
stellar radius, and the parameter $\beta$ is $\sim 1$ for massive
stars. Hence, using the fact that $r^2=z^2+a^2$ and assuming a gas
dominated by protons, we get the particle density of the medium
along the jet axis:
\begin{equation} n(z)=\frac{\dot{M_*}}{4\pi m_p v_{\infty}
(z^2+a^2)}\left(1-\frac{r_*}{\sqrt{z^2+a^2}}\right)^{-\beta}.\label{n(z)}
\end{equation}
Typical mass loss rates and terminal wind velocities for O stars are
of the order of $10^{-5}$ $\dot{M_{\sun}}$ yr$^{-1}$ and 2500 km
s$^{-1}$, respectively (Lamers \& Cassinelli 1999). This simple
modelling for the wind was also used when analyzing the possible TeV
emission from stellar systems (Romero \& Torres 2003, Torres et al.
2004).

It is important to note that we are considering not that the beam
interacts with the wind in a face-on collision, but instead, that
the wind diffuses into the jet from the side. The wind penetration
into the jet outflow depends on the parameter $\varpi \sim v
R(z)/D$, where $v$ is the corresponding velocity of wind/beam in the
direction of diffusion, $R(z)$ is the radius of the jet at a height
$z$ above the compact object, and $D$ is the diffusion coefficient.
$\varpi$ measures the ratio between the diffusive and the convective
timescale of the particles. In the Bohm limit, with typical magnetic
fields $B_0\sim 1-10$ G, $\varpi \leq 1$, and the wind matter
penetrates the jet.

\subsection{Gamma-ray and neutrino emission}

The dominant $\pi$-producing channels in the hadronic interactions
of the jet with the wind are ($E_{\rm th}\sim m_{\pi} c^2$):
\begin{eqnarray}
p+p &\rightarrow& p+\pi^0 \rightarrow p+2\gamma \label{pi0}\\
p+p &\rightarrow& n+\pi^+\rightarrow p+e^-+\bar{\nu}_e+e^+ +\nu_e+
\nu_{\mu} +\bar{\nu}_{\mu}\label{pi+}\\
p+p &\rightarrow& p+ \pi^+ +\pi^- \nonumber\\
 & & \rightarrow p+e^+ +\nu_e+  +2\bar{\nu}_{\mu}+e^- +\bar{\nu}_e +
2\nu_{\mu} \label{pi-},
\end{eqnarray}
where all symbols have their usual meaning in particle physics.
Processes (\ref{pi+}) and (\ref{pi-}) lead to {\sl in situ} pair
creation and neutrino production. The differential gamma-ray
emissivity from $\pi^0$-decays is:
\begin{equation}
q_{\gamma}(E_{\gamma})= 4 \pi \sigma_{pp}(E_p)
\frac{2Z^{(\alpha)}_{p\rightarrow\pi^0}}{\alpha}\;J_p(E_{\gamma},\theta)
\eta_{\rm A}. \label{q} \end{equation} Here, the parameter
$\eta_{\rm A}$ takes into account the contribution from different
nuclei in the wind and in the jet (for standard composition of
cosmic rays and interstellar medium  $\eta_{\rm A}=1.4-1.5$, Dermer
1986). $J_p(E_{\gamma})$ is the proton flux distribution evaluated
at $E=E_{\gamma}$. The cross section $\sigma_{pp}(E_p)$ for
inelastic $p-p$ interactions at energy $E_p\approx 10 E_{\gamma}$
can be represented above $E_p\approx 10$ GeV by
$\sigma_{pp}(E_p)\approx 30 \times [0.95 + 0.06 \log (E_p/{\rm
GeV})]$ mb.  Finally, $Z^{(\alpha)}_{p\rightarrow\pi^0}$ is the
so-called spectrum-weighted moment of the inclusive cross-section.
Its value for different spectral indices $\alpha$ is given, for
instance, in Table A1 of Drury et al. (1994). Notice that
$q_{\gamma}$ is expressed in ph s$^{-1}$ erg$^{-1}$ when we adopt
CGS units.

The spectral gamma-ray intensity (photons per unit of time per unit
of energy-band) is:
\begin{equation}
I_{\gamma}(E_{\gamma},\theta)=\int_V n(\vec{r'})
q_{\gamma}(\vec{r'}) d^3\vec{r'}, \label{I}
\end{equation}
where $V$ is the interaction volume.
The spectral energy distribution is
$L^{\pi^0}_{\gamma}(E_{\gamma},\theta)=E_{\gamma}^2
I_{\gamma}(E_{\gamma},\theta)$ and using eqs. (\ref{Jp_lab}),
(\ref{n01}), (\ref{n02}), (\ref{n(z)}),  (\ref{q}) and (\ref{I}), we
get:
\begin{eqnarray}
&& L^{\pi^0}_{\gamma}(E_{\gamma},\theta)\approx  \;\frac{ q_{\rm j}
z_0^{\epsilon (n-2)} Z^{(\alpha)}_{p\rightarrow\pi^0}}{ 2\pi m_p^2
v_{\infty}} \;\frac{\alpha-1}{\alpha}\;({E'}_p^{\rm min})^{\alpha-1}
\times \nonumber \\ && \dot{M}_* \dot{M}_{\rm disk} \;
\sigma_{pp}(10\; E_{\gamma})\;
 \frac{\Gamma^{-\alpha+1} \left(E_\gamma-\beta_{\rm b}
\sqrt{E_\gamma^2-m_p^2c^4} \cos \theta\right)^{-\alpha}}{\left[\sin
^2 \theta + \Gamma^2 \left( \cos \theta - \frac{\beta_{\rm b}
E_\gamma}{\sqrt{E_\gamma^2-m_p^2 c^4}}\right)\right]^{1/2}}
\nonumber \\ && \int_{z_0}^{\infty} \frac{z^{\epsilon
(2-n)}}{(z^2+a^2)}
\left(1-\frac{r_*}{\sqrt{z^2+a^2}}\right)^{-\beta} dz.
%
%L^{\pi^0}_{\gamma}(E_{\gamma})&\approx& 0.75 \;\frac{ q_{\rm j}
%z_0^{\epsilon (n-2)}
%Z^{(\alpha)}_{p\rightarrow\pi^0}}{ \pi  m_p^2 v_{\infty}}
%\;\frac{\alpha-1}{\alpha}\; \Gamma^{-\alpha}
%(E_p^{\rm min})^{\alpha-1} \times \nonumber \\ && \dot{M}_* \dot{M}_{\rm
%disk} \;
%\sigma_{pp}(10\; E_{\gamma})\; \frac{\left(E_{\gamma}-\beta_{\rm b} \sqrt{E_{\gamma}^2-m_p^2c^4}\right)^{-\alpha}}{1-\frac{\beta_{\rm b} E_{\gamma}}{\sqrt{E_{\gamma}^2-m_p^2 c^4}}} \times \nonumber \\
%&&
%\int_{z_0}^{\infty}
%\frac{z^{\epsilon (2-n)}}{(z^2+a^2)}
%\left(1-\frac{r_*}{\sqrt{z^2+a^2}}\right)^{-\beta} dz.
\end{eqnarray}
This expression gives approximately the $\pi^0$-decay gamma-ray
luminosity for a windy microquasar at energies $E_{\gamma}>1$ GeV,
in a given direction $\theta$ with respect to the jet axis. The
neutrino spectrum roughly satisfies (e.g., Dar \& Laor 1997): $
{dF_{\nu}}/{dE_{\nu}}\simeq 0.7 {dF_{\gamma}}/{dE_{\gamma}}. $

\section{Discussion on neutrino upper limits}
\label{sect:res}
%\hspace{15pt}%                   %% preserved for Editor

Upper limits imposed by AMANDA II (Ahrens et al. 2003) in known
microquasars are typically $F < 4 \times 10^{-7} (E/{\rm GeV})^{-2}
{\rm GeV}^{-1} {\rm cm}^{-2} {\rm s}^{-1}$. This upper limit and the
theoretical expectation for the neutrino production, given
assumptions on the free parameters of the system, allows already to
put some constraints on the most favorable hadronic models. For
instance, the model in which the distance is $D=5$ kpc or less,
$q_j$=10$^{-2}$ or higher, the stellar mass loss rate is 10$^{-5}$
M$_\odot$ yr$^{-1}$ or higher, and the jet is inclined only 10
degrees or less, is already ruled out by neutrino astronomy. Note
that this model maximizes the gamma-ray and neutrino emission, by
enhancing the target matter density for proton interactions, by
locating the microquasar at roughly half the distance to the
Galactic Center, and by assuming a matter content in the jet with a
significant dose of hadrons. In any case, this is a strong result in
the sense that there is no need for any pointing of any telescope
(the only assumption is that the microquasar is in the northern
hemisphere, in the field of view of the AMANDA-II experiment). Note,
however, that still there are plenty of models (different choices of
system parameters) between the atmospheric neutrino background and
the typically imposed upper limits; these have improved
observational expectations for ICECUBE (as an example, models with
proton slope 2.2 are shown in Figure 1).

Note that alternative hadronic microquasar models make use of
photopion production of neutrinos (Levinson \& Waxman 2001). The
underlying idea in this kind of scenarios is that neutrinos are the
result of photopion processes with synchrotron photons and external
target fields surrounding the microquasar jet. Distefano et al.
(2002), have presented the expectations for the neutrino fluxes from
known microquasars under the assumed validity of this model, and
focused on their scrutiny with the forthcoming ICECUBE neutrino
telescope.
%"Neutrino flux predictions for known Galactic microquasars"
Apart from a technical point --ICECUBE search bin is most likely
going to be 1 degree, not 0.3, what would enhance the background
significantly from Distefano et al. estimates, making detection more
difficult-- the strong claim in their study is that they predict in
some cases even hundreds of events per year in a km-scale detector
like ICECUBE. These would make of microquasars the most notable
neutrino sources in the sky. For instance, in the case of SS433, the
prediction is 252 muons events/yr above 1 TeV. But this implies
around 20 or more events per year in AMANDA-II, what is ruled out
already by data. Table 1 shows a comparison between the predictions
of Distefano et al. (2002) and the measured upper limits extracted
from the work of Ahrens et al. (2004), assuming a $E^{-2}$ spectrum,
for some of the cases that appear to be already ruled out, or are on
the verge of being ruled out, by AMANDA-II data. Note that the case
of SS433 is especially significant: in this case it is known for
sure that in the microquasar's jets there are protons and heavy
nuclei accelerated to relativistic speeds. For SS433, the
estimations of Distefano et al. are ruled out by about one order of
magnitude. Further analysis of neutrino upper limits will be
presented elsewhere.

\begin{figure}
   \vspace{2mm}
   \begin{center}
   \hspace{3mm}\psfig{figure=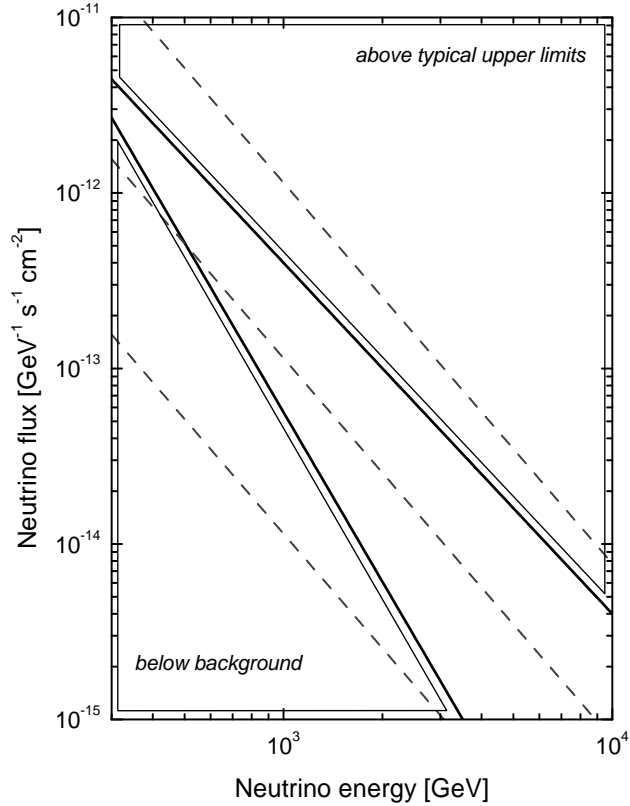,width=80mm,height=110mm,angle=0.0}
%   \parbox{180mm}{{\vspace{2mm} }}
   \caption{Fluxes of neutrinos produced in the framework of different models of
   hadronic microquasars as compared with the atmospheric neutrino background and
   the upper limits for typical sources obtained by AMANDA-II.   }
   \label{neu1}
   \end{center}
\end{figure}

\begin{table}
   \begin{center}
\caption{Comparison between upper limits and predictions of the
photopion model for some microquasars.}
\begin{tabular}{lll}
%\noalign{\medskip}
\hline
%\noalign{\smallskip}
Microquasar & Prediction
  & Upper limit
\\
&  (Distefano et al. 2002)
  &
(data from Ahrens et al. 2003)\\
 & erg cm$^{-2}$
s$^{-1}$
  &  erg cm$^{-2}$
s$^{-1}$ \\
%\noalign{\smallskip}
\hline
%\noalign{\smallskip}
SS433 & 1.7E-9 & 2.6E-10 \\
Cyg X-3 & 4.0E-9 & 1.3E-9\\
Ci Cam & 2.2E-10 & 2.9E-10\\
\hline
\end{tabular}
\label{t1}
   \end{center}
\end{table}

\section{Concluding remarks}

It is clear that, with the appearance of km-scale detectors,
neutrino astronomy will soon become an interesting aid in the study
of microquasars and other stellar objects. Here we have focused on
showing that the current state of the art in neutrino observations
can also be used to establish useful constraints on some of the
hadronic models presented in the literature.

\subsection*{Acknowledgements}

We thank D. Purmohammad for insightful comments regarding the proton
flux treatment. D.F.T. research is done under the auspices of the US
Department of Energy (NNSA), by the UC's LLNL under contract
W-7405-Eng-48. G.E.R. is supported by Fundaci\'on Antorchas, ANPCyT
(PICT 03-04881), and CONICET (PIP 0438/98). This research benefited
from the ECOS French-Argentinian cooperation agreement.

\label{lastpage}

\end{document}